\documentstyle[12pt]{article}
\newlength{\dinwidth}
\newlength{\dinmargin}
\newlength{\extraspace}
\newlength{\extraspaces}
\newcommand{\be}{\begin{equation}
\addtolength{\abovedisplayskip}{\extraspaces}
\addtolength{\belowdisplayskip}{\extraspaces}
\addtolength{\abovedisplayshortskip}{\extraspace}
\addtolength{\belowdisplayshortskip}{\extraspace}}
\newcommand{\ee}{\end{equation}}
\newcommand{\bdm}{\begin{displaymath}
\addtolength{\abovedisplayskip}{\extraspaces}
\addtolength{\belowdisplayskip}{\extraspaces}
\addtolength{\abovedisplayshortskip}{\extraspace}
\addtolength{\belowdisplayshortskip}{\extraspace}}
\newcommand{\edm}{\end{displaymath}}
\renewcommand{\thefootnote}{\fnsymbol{footnote}}
\def\simlt{\mathrel{\lower2.5pt\vbox{\lineskip=0pt\baselineskip=0pt
           \hbox{$<$}\hbox{$\sim$}}}}
\newcommand{\beq}{\begin{equation}}
\newcommand{\eeq}{\end{equation}}
\newcommand{\bea}{\begin{eqnarray}}
\newcommand{\eea}{\end{eqnarray}}
\newcommand{\ts}{\thinspace}
\newcommand{\semi}{; \\}
\newcommand{\pr}{Phys.\ Rev.\ }
\newcommand{\prp}{Phys.\ Rep.\ }
\newcommand{\prl}{Phys.\ Rev.\ Lett.\ }
\newcommand{\np}{Nucl.\ Phys.\ {\bf B}}
\newcommand{\pl}{Phys.\ Lett.\ {\bf B}}
\newcommand{\rmp}{Reviews of Modern Physics\ }

\newcommand{\peetee}{p_T}
\newcommand{\etoc}{\eta_8}
\newcommand{\etas}{\eta_0}
\newcommand{\rhoc}{\rho_8}
\newcommand{\ttbar}{t{\overline t}}
\newcommand{\qqbar}{q{\overline q}}
\newcommand{\ppbar}{p{\overline p}}
\newcommand{\gev}{\ts \rm GeV \ts}
\newcommand{\pico}{\ts \rm pb \ts}
\newcommand{\glu}{\cal G}
\newcommand{\mrhot}{m_{\rhoc}}
\newcommand{\meta}{m_{\etoc}}
\newcommand{\aqc}{\alpha_s}
\newcommand{\FQ}{F_Q}
\newcommand{\mtt}{M_{\ttbar}}
\newcommand{\shat}{\hat s}
\begin{document}
\begin{titlepage}
\begin{flushright}
UTPT-95-07\\
\end{flushright}
\vspace{24mm}
\begin{center}
\Large{{\bf  A New Physics Source of Hard Gluons \\in Top Quark Production}}
\end{center}
\vspace{5mm}
\begin{center}
B. Holdom\footnote{e-mail address: holdom@utcc.utoronto.ca} and
M.~V.~Ramana\footnote{e-mail address:
ramana@medb.physics.utoronto.ca}\\ {\normalsize\it Department of
Physics}\\ {\normalsize\it University of Toronto}\\ {\normalsize\it Toronto,
Ontario,}\\ {\normalsize CANADA, M5S 1A7}
\end{center}
\vspace{2cm}
\thispagestyle{empty}
\begin{abstract}
We consider the contribution of new strongly interacting sector, with a
characteristic scale of half a TeV, to top quark production at the
Tevatron.  The color-octet, isosinglet analog of the $\rho$ meson in
this theory is produced copiously in hadron colliders. If the mass of
this resonance is less than twice the mass of the lightest
pseudo-Goldstone bosons, then an important decay mode could be to the
color-octet analog of the $\eta$ and a gluon.  The subsequent decay of
this $\eta$ into $\ttbar$ gives rise to top quark events with a hard
gluon.
\end{abstract}
\end{titlepage}
\newpage

\renewcommand{\thefootnote}{\arabic{footnote}}
\setcounter{footnote}{0}
\setcounter{page}{2}

Recently the CDF and D0 collaborations have presented evidence confirming
the production of top quarks at the Tevatron Collider.  CDF \cite{cdfnew}
finds a top quark mass of $m_t = 176 \pm 8 \pm {10} \gev$ and the cross
section $\sigma(\ppbar \to \ttbar) = 6.8^{+3.6}_{-2.4} \pico$. D0 \cite{d0}
finds a mass of $m_t = 199^{+19}_{-21}\pm 22 \gev$ and a cross section $6.4
\pm 2.2 \pico$. For $m_t=176 \gev$ the QCD prediction for the cross section
is $\sigma(\ppbar \to \ttbar) = 4.79^{+0.67}_{-0.41} \pico$
\cite{laenen}.  While the CDF value for the cross section is not
inconsistent with the QCD prediction, it still leaves some room for
contributions to $\ttbar$ production from other sources.  It is of
interest to explore the new physics signals accessible to more
detailed analyses of top quark production in the future.

In this paper we will consider the contribution of a color-octet, isosinglet
vector resonance, the $\rhoc$, to $\ttbar$ production. Such a resonance is
expected in any strongly interacting theory which has at least one colored,
electroweak doublet of fermions. For the most part we consider $\rhoc$
masses in the 400-600 GeV range, although we shall see that higher values
may also be of interest.  Examples of theories with a characteristic scale
somewhat below the usual TeV electroweak symmetry breaking scale are
discussed in \cite{multi}, \cite{first}, and \cite{meta}.

As discussed in \cite{bob}, the $\rhoc$ has the odd-parity decay modes
$\rhoc \to \etoc \glu$ and $\rhoc \to \etas \glu$.  If the color-octet or
the color-singlet analogs of the QCD $\eta$ meson ($\etoc$ or $\etas$) are
above the $\ttbar$ threshold, then they decay predominantly into $\ttbar$
\cite{drk}. A signature to be explored in this paper is provided by the hard
gluon produced in addition to the $\ttbar$ pair.

The color-octet $\etoc$ itself can be strongly produced in hadronic
collisions and its contribution to $\ttbar$ production has been
considered by several authors \cite{drk} -- \cite{ken}.  However this
is mainly produced from initial states involving two gluons. At the
Tevatron, since we are dealing with processes where the initial
partons have a large fraction of the proton's momentum, the gluon
distribution functions are much smaller than the quark distribution
functions. This leads us to consider the $\rhoc$ which is produced
from $\qqbar$ initial states.

If the $\rhoc$ is below the two pseudo-Goldstone boson threshold, the decays
$\rhoc \to \etoc \glu$ and $\rhoc \to \etas \glu$ have sizeable branching
fractions for a large range of parameter values. Likewise, if allowed, the
decays $\etoc \to \ttbar$ and $\etas \to \ttbar$ are the dominant modes of
decay \cite{drk} of the $\etoc$ and the $\etas$ (except when their masses are
very close to the $\ttbar$ threshold). The color factors are such that the
contribution a $\etas$ makes to the cross section is two-fifths the
contribution of an $\etoc$ with the same mass.  On the other hand for the
mass ranges of the $\etoc$ that we will be exploring in this paper, the mass
of the $\etas$ may well be below the $\ttbar$ threshold. To be definite we
will henceforth assume that only the $\etoc$ contributes to the process of
interest. Since the production cross section for a color-octet $\rhoc$ is
quite large \cite{ehlq,first}, we expect that the process $$\qqbar \to \rhoc
\to \etoc {\glu} \to \ttbar {\glu}$$ could contribute substantially to the top
quark production rate at the Tevatron, depending on the masses of the $\rhoc$
and the $\etoc$.

The CDF results on the top quark mass and production cross section are
primarily based on detecting the $\ttbar$ pair in its leptons + jets decay
mode. This would nominally produce events with four jets. However
many of the events observed have five or more jets, as could be expected
from QCD corrections. For such events the procedure followed \cite{cdfnew}
is to assume that the four highest $E_T$ jets arise from the decay of the
$t\overline t$ system. The point is that events which have extra gluons are
included in the CDF analysis, and hence our mechanism for the production of
$\ttbar \glu$ would contribute to their measurement of the $\ttbar$ cross
section.

\begin{table}
\centering
\begin{tabular}{|c|c|p{0.38in}|p{0.38in}|p{0.25in}|c|c|c|}\hline
\  $\mrhot$ \ &\ $\meta$ \  &\multicolumn{3}{c|}
{$\sigma({\ppbar \to \rhoc \to \ttbar + \glu})$}&\multicolumn{3}{c|}
{$\sigma({\ppbar \to \etoc \to \ttbar})$ }\\ [1mm] \hline
475 & 400 & 2.06 & 2.74 &2.99& 0.72 & 1.61 & 1.35 \\ [2mm]
475 & 450 & 0.71 & 0.72 &0.60& 0.42 & 0.83 & 0.40 \\ [2mm]
525 & 450 & 1.35 & 1.64 &1.69& 0.35 & 0.69 & 0.33 \\ [2mm]
750 & 400 & 0.25 & 0.41 &0.25& 0.35 & 0.72 & 0.34 \\ [2mm]
\hline
\end{tabular}
\label{mej1}

\caption{Cross sections for the production of $\ttbar + \glu$
at the Tevatron. For each set of masses, the first column under each
cross section corresponds to \{$N=2,k=1.0$\}, the second to
\{$N=4,k=1.0$\} and the third to \{$N=2,k=0.1$\}. All masses are in GeV
and all cross sections are in pb. No QCD corrections are included.}

\end{table}

Before presenting the cross sections for $\ttbar$ production, we first show the
partial widths for the various decay modes that are relevant to our
calculations.  The partial widths for the decay modes\footnote{Assuming that
$\rhoc$ is below the two pseudo-Goldstone boson threshold.} of the $\rhoc$ can
be calculated using the ideas outlined in \cite{vmd,bky,mike} to be:
\beq
\Gamma(\rhoc \to \qqbar) = {5 \over {6}} \ts {\aqc^2(\mrhot) \over
\alpha_{\rhoc}}
\ts \mrhot \ts,
\eeq
\beq
\Gamma(\rhoc \to {\glu \glu}) = {1 \over {2}}
\ts {\aqc^2(\mrhot) \over \alpha_{\rhoc}}
\ts \mrhot \ts,
\eeq
and
\beq
\Gamma(\rhoc \to \etoc {\glu}) = {5 \over {128}}{1 \over{\pi^3}}
{\left(N \over {3}\right)^2} {\alpha_{\rhoc} \aqc(\mrhot) \over {\FQ^2}}
{\left({{\mrhot^2 - \meta^2} \over \mrhot}\right)^3}
\ts
\eeq
Following \cite{ehlq}, we have assumed that the
$\rhoc$ decay constant is
\beq
\alpha_{\rhoc} = {3 \over {N}} \ts 2.97 \ts
\eeq
where 2.97 is the QCD $\rho$ meson decay constant.  $N$ is the number of
``colors'' in this theory; if this were to be a technicolor theory then $N$
would be the number of technicolors. Similarly we use the scaling relation
\beq
{\FQ \over \mrhot} = {f_{\pi} \over {m_{\rho}}} \ts \sqrt{{{N} \over
3}}   \label{scaling}
\eeq
to obtain $\FQ$, the analog of the pion decay constant.

A survey of values of the ratio of $\Gamma(\rho \to \etoc {\glu})$ to
the dijet width ($\Gamma(\rho \to \qqbar) + \Gamma(\rho \to {\glu
\glu})$) shows that it is indeed sizeable for a large range of parameters.
For example, with $N = 2$ and $\mrhot = 475 \gev$, $\meta = 400 \gev$,
the ratio is $0.20$, while for $\mrhot = 525 \gev$, $\meta = 450 \gev$
the ratio is $0.15$.

If $\meta > 2 m_t$ the decay width of the $\etoc \to \ttbar$ is
given by \cite{drk}
\beq
\Gamma (\etoc \to \ttbar) = k_t {{m_t^2 \meta} \over {16 \pi \FQ^2}}
{\sqrt{\left(1-{{4 m_t^2} \over {\meta^2}}\right)}} \ts.
\eeq
The constant of proportionality $k_t$ is determined by some underlying
theory and is expected to be of order one. In our calculations, $k_t$
is varied between 0.1 and 1.  A competing decay mode of the $\etoc$ is to
$\glu \glu$; the decay width for this mode is
\beq
\Gamma (\etoc \to {\glu \glu}) = {{5 N^2 \aqc \meta^3}
\over {384 \pi^3 \FQ^2}} \ts.
\eeq
The decay $\etoc \to \ttbar$ dominates over the decay into $\glu \glu$
except when $\meta$ is very close to the $\ttbar$ threshold or when
$k_t$ is very small.

\begin{figure}
\input{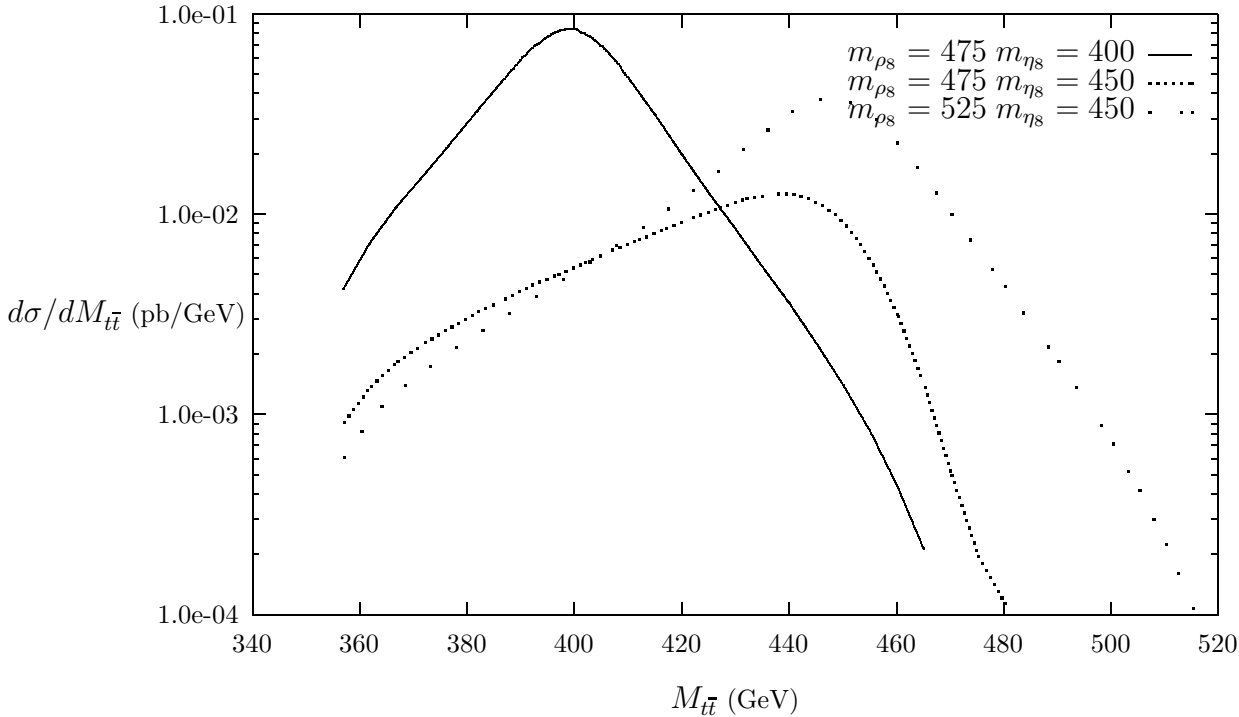}
\vspace{ 0.1 in}
\caption{The differential cross section for total $\ttbar + \glu$
production with $\mtt$ being the invariant mass of $\ttbar$.  No
rapidity cuts are imposed. $N = 2$ and $k_t = 1$.}
\label{dcsy92}
\end{figure}

The resulting partonic level differential cross section for the process
$\qqbar \to \ttbar {\glu}$ is
\bea
{{d{\hat \sigma}} \over {dz d\mtt}} &=& {5 \over {384 \pi^4}} \ts
{\left({N \over 3}\right)^2} \ts {{\aqc^3 k_t m_t^2} \over \FQ^4} \ts
{\mrhot^4 \over {((\shat-\mrhot^2)^2 + \shat \Gamma^2_{\rhoc})}}
\nonumber \\
& &\left(1 - {\mtt^2 \over \shat}\right)^2
\sqrt{\left(1-{{4 m_t^2} \over {\mtt^2}}\right)} \ts
{{(\mtt^2 - 2m_t^2) \mtt } \over {((\mtt^2 - \meta^2)^2 + \mtt^2
\Gamma^2_{\etoc})}} \ts
(1+ z^2) \ts.
\eea
where $\mtt$ is the invariant mass of the $\ttbar$ pair and $z = {\rm
cos}(\theta)$, where $\theta$ is the angle between the outgoing gluon and the
incoming quark. The partonic center of mass energy is $\shat$. We see that
the gluons produced are mildly peaked in the forward and backward
directions.  In view of the complexity of the top quark analysis followed by
the CDF collaboration \cite{cdf}, no rapidity cuts are imposed on the top
quarks. We also do not include any QCD corrections although they are
expected to be significant, as in the case of $\ttbar$ production from
QCD \cite{laenen}.  These corrections could easily increase the production
rate by 50\% or more.

This parton level result can be folded in with the quark distribution
functions and integrated over $z ={\rm cos}(\theta)$ and $\mtt$ in order to
obtain the total production cross section for $\ttbar + {\glu}$.  We have
ignored the contribution from ${\glu \glu} \to \ttbar + \glu$ since the gluon
structure functions are small at the Tevatron. All the results in this work
use the EHLQ set II structure functions \cite{ehlq}\ with $Q^2 = \hat{s}$.

The results of this computation are presented in Table~1 which shows the
cross sections for $\ttbar$ production for different values of $\mrhot$,
$\meta$, $N$ and $k_t$. We have chosen values of $\mrhot$ consistent with the
range excluded in \cite{dijet}. That analysis assumes that the $\rhoc$ decays
only to dijets --- this is not strictly valid for the vector particles
considered here since they have other decay modes open to them. Nevertheless
the results in \cite{dijet} indicate that it is unlikely that there are
colored resonances in the $\approx 200-400 \gev$ mass range.

The total cross section for $\ttbar$ production corresponds to events
where the extra (hard) gluon may or may not be detected. For
comparison, along with the cross sections for $\ppbar \to \rhoc \to
\ttbar + \glu$, we also show the contribution from $\ppbar \to \etoc
\to \ttbar$ using the formulae in \cite{ken}.  However, unlike the
authors of \cite{ken}, we do not incorporate any estimates of QCD
corrections.  Also, given a value of $\mrhot$ we use
Eq.~(\ref{scaling}) to determine the value of $\FQ$. Thus the values
given in Table~1 for $\sigma(\ppbar \to \etoc \to \ttbar)$
depend on our choice of $\mrhot$.

The largest $\rhoc$ mass displayed in Table~1 is 750 GeV; this mass is
beginning to approach a value more typical of standard technicolor
theories. In this case, the contribution to top quark production from
our mechanism is fairly small (about 5\% or less of the estimated
cross section). However these events would have a unique signature ---
a $\ttbar$ pair with large $\peetee$ recoiling against a high
energy gluon.

\begin{figure}
\input{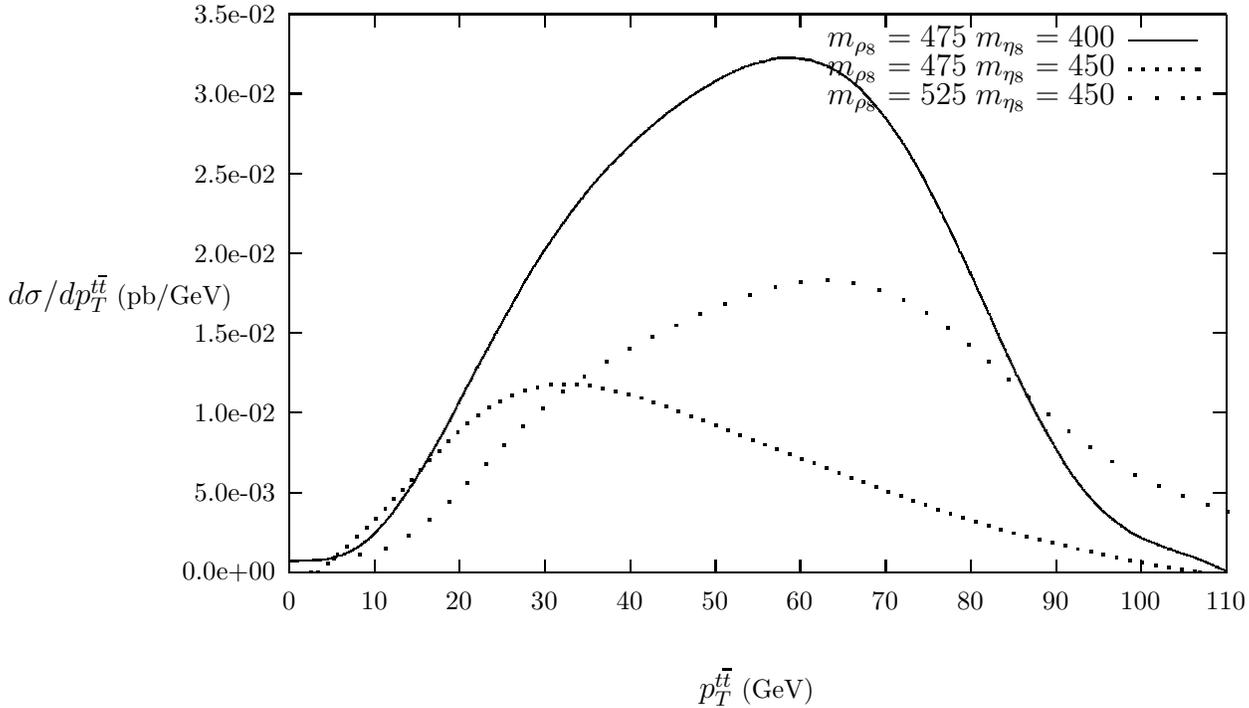}
\vspace{ 0.1 in}
\caption{The $p_T^{\ttbar}$ distribution in $\ttbar + \glu$
production.  A rapidity cut of $y_c = 2.0$ is imposed on the final
state gluon. $N = 2$ and $k_t = 1$.}
\label{gluet}
\end{figure}

For the lower $\mrhot$ values we consider in more detail methods
to distinguish our contribution to top quark production from the
standard model contribution. In order to do this we focus on three
different distributions --- the invariant mass $\mtt$ and the
transverse momentum $p_T^{\ttbar}$ of the $\ttbar$ pair, and the total
invariant mass $M_{inv}$ of $\ttbar +\glu$. These distributions are
plotted Figures (1-3) for the parameter values : $\{\mrhot = 475 \gev$,
$\meta = 400 \gev\}$, $\{\mrhot = 475 \gev$, $\meta = 450 \gev\}$ and
$\{\mrhot = 525 \gev$, $\meta = 450 \gev\}$.

The differential cross section ${{d{\hat \sigma}}/{d\mtt}}$ is displayed in
Fig.~\ref{dcsy92}. Since we are interested in estimating the total top quark
production, we have not imposed any rapidity cuts on the gluon.  We notice
the peak in the $\mtt$ distribution due to the intermediate resonance, the
$\etoc$. However for heavier $\etoc$, the $\mtt$ distribution is quite
broad and the peak may not show up very cleanly.

\begin{figure}
\input{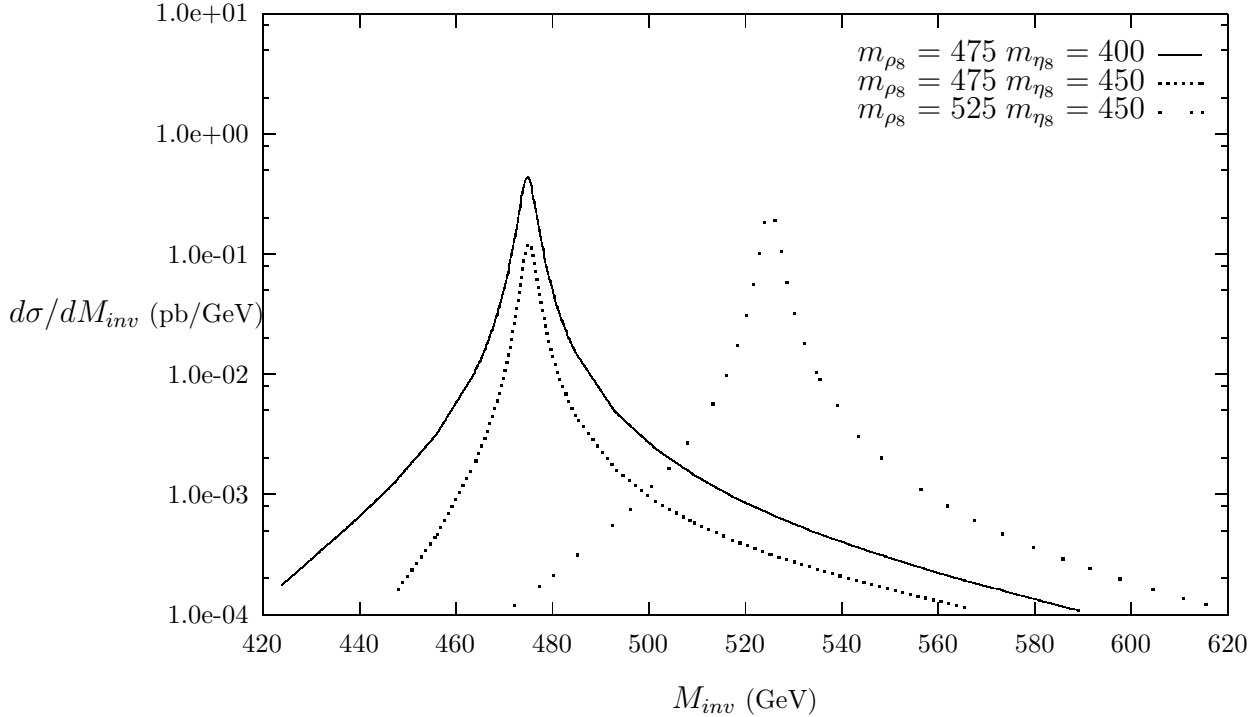}
\vspace{ 0.1 in}
\caption{The differential cross section for $\ttbar + \glu$
production with $M_{inv}$ being the invariant mass of $\ttbar + \glu$.
A rapidity cut of $y_c = 2.0$ is imposed on the final state gluon.  $N
= 2$ and $k_t = 1$.}
\label{dc3}
\end{figure}

Another variable of interest for new physics is the transverse momentum
of the $\ttbar$ pair $p_T^{\ttbar}$.  We show the distribution from our
mechanism in Fig.~\ref{gluet}. In calculating this distribution (and
the one in Fig.~\ref{dc3}) we impose a rapidity cut of $y_c = 2.0$, so
that the gluon be within the ambit of the detector.  This is the value
used by the CDF collaboration \cite{cdf}; we found that this cut
reduced the total event rates by about ten to fifteen percent. The
standard model $p_T^{\ttbar}$ distribution has been obtained
\cite{nason} using both next-to-leading order QCD calculations and
HERWIG simulations. The distribution peaks for $p_T^{\ttbar}$ around 5
GeV and then falls rapidly for increasing $p_T^{\ttbar}$. A significant
excess of high $p_T$ $\ttbar$ pairs would signal a nonstandard
production mechanism.

The $p_T^{\ttbar}$ distribution is closely related to the distribution
of the quantity referred to as $X$ in \cite{cdf}.  This is the total
{\it observed} transverse energy remaining in the event after
subtracting vectorially the transverse energy of the four jets and the
charged lepton.  The $X$ values for the seven events used for mass
fitting in \cite{cdf} are \{7.1~GeV, 7.7~GeV, 14.8~GeV, 17.0~GeV,
20.0~GeV, 26.3~GeV and 35.6~GeV\}.  A larger data set is clearly required
before any serious analysis can be made.

We notice that there is a significant bias in the extraction of $X$ which
distorts its interpretation as a measure of $p_T^{\ttbar}$. The method
\cite{cdfnew} used in deciding which of the jets in an event come from top
decays is to assign the four highest energy jets to the decay of the
$\ttbar$ system.  While this is suitable for standard model $\ttbar$
production, it may not be appropriate if one wishes to allow for new
physics. In particular, the hard gluon produced in our mechanism can be
mistakenly assigned as one of the jets from the $\ttbar$ system.  This in
turn could distort the extraction of the top mass.  It also implies that the
observed $X$ distribution is likely to be softer than the true
$p_T^{\ttbar}$ distribution.

The third variable we consider is the total invariant mass of the event,
which in our case is the invariant mass of the $\ttbar +\glu$.  Since the
$\rhoc$ is relatively narrow this distribution is sharply peaked, as can
be seen in Fig.~\ref{dc3}. This variable is also less prone to
experimental ambiguities involved in assigning jets to underlying partons.

In conclusion, we have suggested a mechanism for the production of top
quarks at the Tevatron. The $\rhoc$ and $\etoc$ resonances which take
part in this resonant enhancement of $\ttbar$ events are generic of
any strongly interacting theory with at least one colored, electroweak
doublet of fermions. If these resonances have masses in the range of
400-550~GeV range, they can contribute significantly to $\ttbar$
production at the Tevatron.  We have suggested ways of distinguishing this
mechanism from standard model production.  We finally note that if the
color-singlet $\etas$ was also above the top-pair threshold then we
would have contributions from both the $\etoc$ and the $\etas$. There would
be two peaks in the $M_{\ttbar}$ distribution, with their relative
contributions being fixed by the masses of the resonances.

{\bf Acknowledgements}

We would like to thank P.~Sinervo and G.~Triantaphyllou for useful
discussions.  This research was supported in part by the Natural
Sciences and Engineering Research Council of Canada.


\newpage
\listoffigures

\end{document}